\documentclass{aa}

\usepackage{natbib}
\bibpunct{(}{)}{;}{a}{}{,} 

\usepackage{graphicx,hyperref,url}
\usepackage[normalem]{ulem}
\usepackage{mathrsfs,amssymb,amsmath}
\usepackage[usenames]{color}
\usepackage{subfigure}

\usepackage[varg]{txfonts}

\newcommand{\ud}{\mathrm{d}}
\newcommand{\be}{\begin{equation}}
\newcommand{\ee}{\end{equation}}

\def\HII{H{\sc ii} }

\newcommand*\xbar[1]{%
   \hbox{%
     \vbox{%
       \hrule height 0.5pt 
       \kern0.2ex
       \hbox{%
         \kern-0.1em
         \ensuremath{#1}%
         \kern-0.0em
       }%
     }%
   }%
}

\begin{document}

\title{Non-thermal emission from cosmic rays accelerated in \HII regions}

\author{Marco~Padovani\inst{1}, Alexandre~Marcowith\inst{2},
\'Alvaro S\'anchez-Monge\inst{3}, Fanyi Meng\inst{3}, Peter Schilke\inst{3}}

\authorrunning{M. Padovani et al.}


\institute{INAF--Osservatorio Astrofisico di Arcetri, Largo E. Fermi 5, 50125
Firenze, Italy\\
\email{padovani@arcetri.astro.it}
\and
Laboratoire Univers et Particules de Montpellier, UMR 5299 du CNRS, Universit\'e de Montpellier, place E. Bataillon, cc072, 34095 Montpellier, France
\and
I. Physikalisches Institut, Universit\"at zu K\"oln, Z\"ulpicher Str. 77, 50937 K\"oln, Germany
}


\abstract 
{Radio observations at metre-centimetre wavelengths
shed light on the nature of the emission of \HII regions. Usually this category of objects is dominated by thermal radiation produced by ionised hydrogen, namely protons and electrons. However, a number of observational studies have revealed the existence of \HII regions with a mixture of thermal and non-thermal radiation. The latter represents a clue as to the presence of relativistic
electrons.\ However, 
neither the interstellar cosmic-ray electron flux
nor the flux of secondary electrons, produced by primary cosmic rays
through ionisation processes, is high enough to explain the observed
flux densities.}
{We investigate the possibility of accelerating local thermal electrons
up to relativistic energies in \HII region shocks.}
{We assumed that relativistic electrons can be accelerated through
the first-order Fermi acceleration mechanism and 
we estimated the emerging electron fluxes, the corresponding flux 
densities, and the spectral
indexes.}
{We find flux densities of the same order of magnitude 
of those observed.
In particular, we applied our model to the `deep south' (DS) region
of Sagittarius~B2 and we succeeded in reproducing the observed flux
densities with an accuracy of less than 20\% as well as the spectral
indexes. The model also gives constraints on 
magnetic field strength ($0.3-4$~mG), 
density ($1-9\times10^4$~cm$^{-3}$), 
and flow velocity in the shock reference frame ($33-50$~km~s$^{-1}$) 
expected in DS.}
{We suggest a mechanism able to accelerate thermal electrons inside
\HII regions through the first-order Fermi acceleration.
The existence of a local source of relativistic electrons
can explain the origin of 
both the observed non-thermal emission and the corresponding
spectral indexes.
}


\maketitle

\section{Introduction}
\label{intro}
Massive stars ($M\gtrsim8~{\rm M}_\odot$)
strongly affect the dynamics, the morphology, and the chemistry of their hosting molecular clouds and galaxies, especially during the first stages of formation till their death as supernovae. 
Their scarcity, short lives, and large distances make them difficult to observe, and also because they are deeply embedded in their parental cloud at least during about 15\% of their lifetimes 
\citep{ch02}.
However, even during the embedded phase, massive stars announce their presence by expanding \HII regions. 
This class of objects is routinely observed at infrared and radio wavelengths; the dust close to the \HII region absorbs most of the radiation from the star, emitting in the mid- and far-infrared, while the cool dust farther from the star absorbs the infrared radiation, emitting at sub-millimetre wavelengths. At metre-centimetre wavelengths, the emission is dominated by the interaction of free electrons and protons within the ionised gas.

Multiwavelength studies of \HII regions represent the classic approach to gather information on their morphology \citep{wc89} and their evolutionary stage \citep{ku05, sb13}.
Observations at metre-centimetre wavelengths
allow us to compute the spectral energy distribution, $S_\nu\propto\nu^\alpha$, and discriminate
between thermal ($-0.1<\alpha<2$) 
and non-thermal emission ($\alpha<-0.1$). 
While the former can have different origins,
such as (non-)homogeneous \HII regions; ionised equatorial winds; 
photoevaporated disc winds; thermal radiojets; dense interstellar shock waves
\citep[see][and references therein]{sk13}, the latter only originates
from young stars with active magnetospheres giving rise to gyro-synchrotron emission \citep[e.g.][]{fm99}; and from
fast shocks in discs or jets \citep[e.g.][]{ra95,ss04,sm19}
by the first-order Fermi acceleration mechanism \citep[e.g.][]{cw90,ph15,pm16,rc17}.
\citet{rm93} showed that a negative spectral index could also be due to dust
absorption, but this would require dust mass 
column densities larger
than $10^3$~g~cm$^{-2}$. 

\HII regions are usually dominated by thermal emission \citep[e.g.][]{wc89, ku05, sm2008, sm2011, hoare2012, purcell2013, wang2018, yang2019}. However, observations with sensitive facilities such as the Very Large Array (VLA) or the Giant Metrewave Radio Telescope (GMRT) have revealed the presence of non-thermal emission in a handful of objects. Non-thermal emission in \HII regions
usually shows up 
as spots surrounded by thermal emission 
\citep[e.g.][]{nv16,vv16}, contiguous to thermal emission
as in cometary \HII regions \citep[e.g.][]{mk02},
or sometimes it appears  isolated \citep{ms19}.
Non-thermal emission observed in \HII regions is the fingerprint
of the presence
of relativistic electrons, but they cannot have interstellar origin.
In fact, the interstellar cosmic-ray electron
flux based on the most recent Voyager~1 observations 
\citep{cs16}
is too low to explain the observed flux density.
The same conclusion 
holds for the flux of secondary electrons created through ionisation processes by interstellar cosmic rays (protons and electrons)
whose flux is strongly attenuated
since \HII regions are located in embedded parts of molecular 
clouds (\citealt{pgg09,pi18}; see also Sect.~\ref{model}).

A possible solution to explain the origin of 
these relativistic electrons is through a local acceleration inside
the \HII region itself.
Several mechanisms can be responsible for particle acceleration,
such as 
turbulent second-order Fermi acceleration \citep{pra11}, 
acceleration by shocked background turbulence \citep{gj07},
non-relativistic shear flow acceleration \citep{rd06},
and acceleration in magnetic reconnection sites \citep{dgdp05}.
In this paper we focus on the first-order Fermi acceleration mechanism,
also known as diffusive shock acceleration, according to which charged particles
gain energy while crossing a shock back and forth due to the presence of
magnetic fluctuations around the shock. Thus, charges are extracted from
the thermal pool, accelerate up to relativistic 
energies, and escape in the downstream medium \citep[e.g.][]{dru83,kirk94}.

This paper is organised as follows.
In Sect.~\ref{model} we describe the model for particle acceleration
and we compute the expected flux of shock-accelerated electrons in \HII
regions, in Sect.~\ref{synchtrotonemission} we recall the basic equations
to compute the synchrotron specific emissivity and the corresponding
flux density expected in \HII regions,
in Sect.~\ref{applications} we use our model to 
explain the origin of non-thermal emission in Sagittarius~B2, and
in Sect.~\ref{conclusions} we discuss the implications of our results
and summarise our most important findings.


\section{Model for thermal particle acceleration}
\label{model}

In this section we recall the main equations to compute the timescales involved in particle acceleration at a shock surface, computing the maximum energies and the emerging fluxes of shock-accelerated protons and electrons. For a detailed review of the methods, see \citealt{dd96} and \citealt{ph15,pm16}.

\subsection{Timescales}\label{timescales}
Thermal protons have to be accelerated before $(i)$ they lose energy from collisions, $(ii)$ they diffuse towards the source, and $(iii)$ the shock disappears.
The acceleration timescale, $t_{\rm acc}$, is given by
\be\label{tacc}
t_{\rm acc}=2.9(\gamma-1)\frac{r[1+r(k_{\rm d}/k_{\rm u})^{\sigma}]}{k_{\rm u}^{\sigma}(r-1)} U_{2}^{-2}B_{-5}^{-1}~\mathrm{yr}\,,
\ee
where $U_{2}$ and $B_{-5}$ are the upstream flow velocity in the shock reference frame in unit of 100~km~s$^{-1}$ and the upstream magnetic field strength in unit of 10~$\mu$G, respectively.\ Furthermore, $k_{\rm u,d}$ is the upstream and 
downstream diffusion coefficient that is normalised to the Bohm coefficient
\be\label{diffcoeff}
k_{\rm u}=\left(\frac{\kappa_{\rm u}}{\kappa_{\rm B}}\right)^{-\sigma}=\left(\frac{3eB}{\gamma\beta^{2}m_pc^{3}}\kappa_{\rm u}\right)^{-\sigma}
\ee
with $e$ the elementary charge, $\gamma$ the Lorentz factor,  $\beta=\gamma^{-1}\sqrt{\gamma^{2}-1}$, $m_p$ the proton mass, and $c$ the light speed. 
For a perpendicular shock $k_{\rm u}=rk_{\rm d}$ and $\sigma=1$, while for a parallel\footnote{A parallel and perpendicular shock is when the shock normal is parallel and perpendicular, respectively, to the ambient magnetic field.} shock $k_{\rm u}=k_{\rm d}$ and $\sigma=-1$.
Here, $r$ is the shock compression ratio defined by
\be\label{compressionratio}
r=\frac{\left(\gamma_{\rm ad}+1\right)M_{s}^{2}}{\left(\gamma_{\rm ad}-1\right)M_{s}^{2}+2}\,,
\ee
where
\be\label{Mson}
M_{s}=\frac{U}{c_{s}}
\ee
is the sonic Mach number,
\be\label{cs} 
c_{\rm s}=9.1[\gamma_{\rm ad}(1+x)T_{4}]^{0.5}~\mathrm{km~s^{-1}}
\ee
is the sound speed, $x=n_i/(n_n+n_i)$ is the ionisation fraction ($n_i$ and $n_n$ are the ion and neutral volume density, respectively), and $T_{4}$ is the upstream temperature in unit of $10^{4}$~K. 
In the following, the adiabatic index is set to $\gamma_{\rm ad}=5/3$.
Also, we assume Bohm diffusion ($k_{\rm u}=1$) and the shock to be parallel. For the case 
of deviations from Bohm diffusion regime and perpendicular shocks, see Sect.~\ref{nonbohm}.

The general equation for the collisional energy loss timescale is given by
\be\label{tloss}
t_{\rm loss}=10\frac{\gamma-1}{\beta} n_{6}^{-1} L_{-16} ~\rm{yr} \ ,
\ee
where $n_{6}$ is the volume density in unit of $10^{6}$~cm$^{-3}$ and $L_{-16}$ is the energy loss function in unit of $10^{-16}$~eV~cm$^{2}$ \citep[see Sect. 3 and Fig. 1 in][]{pi18}. We can evaluate the mean loss timescale by averaging it over the particle's up- and downstream residence times \citep{Parizot06} 
%

\be\label{tloss1}
\langle t_{\rm loss}\rangle= \left(\frac{t_{\rm loss,u}^{-1}+r t_{\rm loss, d}^{-1}}{1+r}\right)^{-1}\ .
\ee
The up- and downstream loss timescales differ in the density (downstream is a factor $r$ higher) and in the Coulomb component of the energy loss function, which depends on temperature\footnote{The relation between up- and downstream temperatures is given by the classic Rankine-Hugoniot condition.} \citep{ms94}.

The upstream diffusion timescale, $t_{\rm diff,u}$, is obtained by assuming that the diffusion length in the upstream medium has to be a fraction $\epsilon<1$ of the distance between the central source and the shock, $R$, which is $\kappa_{\rm u}/U=\epsilon R$. The corresponding timescale is given by
\be\label{tescu}
t_{\rm diff,u}=22.9\epsilon\frac{k_{\rm u}^{\sigma}}{\gamma\beta^{2}}B_{-5}R_2^2~\mathrm{yr}\,,
\ee
where $R_2$ is the shock radius in unit of 100~AU. In the following we assume $\epsilon=0.1$. Since we modelled the particle acceleration in a \HII region, there is no further condition on the downstream escape timescale such as in protostellar jet shocks \citep[see Sect. 2.5 in][]{pm16}.

The determination of the age 
of an \HII region is not straightforward and it can be estimated in the following different ways: through `chemical clocks' \citep[e.g.][]{trevinomorales2014}, statistics \citep[e.g.][]{wc89}, and simulations \citep[e.g.][]{peters2010a}. All these methods give a lifetime of the order of $\approx10^4$--$10^6$~yr. Here, we computed the dynamical timescale of an \HII region, $t_{\rm dyn}$, as a function of the shock radius and velocity,
\be\label{tdyn}
t_{\rm dyn}=4.7R_2U_2^{-1}~\mathrm{yr}\,.
\ee
\subsection{Non-Bohm diffusion regimes
and perpendicular shocks}
\label{nonbohm}


The upstream diffusion coefficient can be written
as a function of the magnetic field strength and its
turbulent component $\delta B$ \citep{dru83}
\be
k_{\rm u}=\left(\frac{\kappa_{\rm u}}{\kappa_{\rm B}}\right)^{-\sigma}=\left(\frac{B}{\delta B}\right)^2\,.
\ee
The hypothesis of the Bohm diffusion regime implies $\delta B=B$, namely the magnetic fluctuations that determine the pitch angle scattering are large and particle acceleration is effective and works at its maximum degree. In the case of parallel shocks, deviations from the Bohm regime 
($k_{\rm u}>1$)
correspond to a reduction of $\delta B$, resulting in a lower acceleration efficiency and a decrease of the maximum energy reached, $E_{\rm max}$.
Besides, the acceleration timescale increases
since 
$t_{\rm acc}\propto k_{\rm u}$ (Eq.~\ref{tacc})
and the parameter space where particle
acceleration is efficient is reduced because
of Coulomb losses (see Sect.~\ref{sectEmaxp}
and Fig.~\ref{plottimescales} for more details).

For perpendicular shocks, the acceleration timescale decreases by a factor $(r+1)/2$ with respect to the parallel case and this results in a slight increase of $E_{\rm max}$. However, in the case of perpendicular transport, $E_{\rm max}$ is reduced by magnetic field line wandering \citep{kd96} and $k_{\rm u}$ is limited by requiring that particles have to be scattered in the time required to drift through the shock in order to avoid any anisotropy in the distribution of the accelerated particles \citep{jok87}. For further details on perpendicular shocks and the non-Bohm regime, see Sect.~5.4.1 in \citet{pm16}.

\subsection{Maximum energy and emerging flux at the shock surface}\label{Emaxshock}

In the following subsections we describe how the maximum energy and the shock-accelerated particle flux is obtained for thermal protons (Sect.~\ref{sectEmaxp}) and electrons (Sect.~\ref{secEmaxe}).

\subsubsection{Acceleration of thermal protons}\label{sectEmaxp}
In order to provide an efficient acceleration, the flow has to be
supersonic and super-Alfv\'enic, namely
\be\label{maxcsva}
U>\max(c_{\rm s},c_{\rm A})\,,
\ee
where the Alfv\'en speed is given by
\be\label{alfvenspeed}
c_{\rm A}=2.2\times10^{-2}n_6^{-0.5}B_{-5}~\mathrm{km~s^{-1}}\,.
\ee
If this condition is fulfilled, the maximum energy reached by a thermal proton, $E_{\rm max,p}$, imposes 
\be\label{Emaxcond}
t_{\rm acc}=\min(t_{\rm loss},t_{\rm diff,u},t_{\rm dyn})\,.
\ee
We note that we assume the medium to be completely ionised ($x=1$)
as expected in \HII regions. In Appendix~\ref{app:xneq1} we show that as soon as $x<1,$ the particle
acceleration efficiency drops.

\begin{figure}[!t]
\begin{center}
\resizebox{\hsize}{!}{\includegraphics{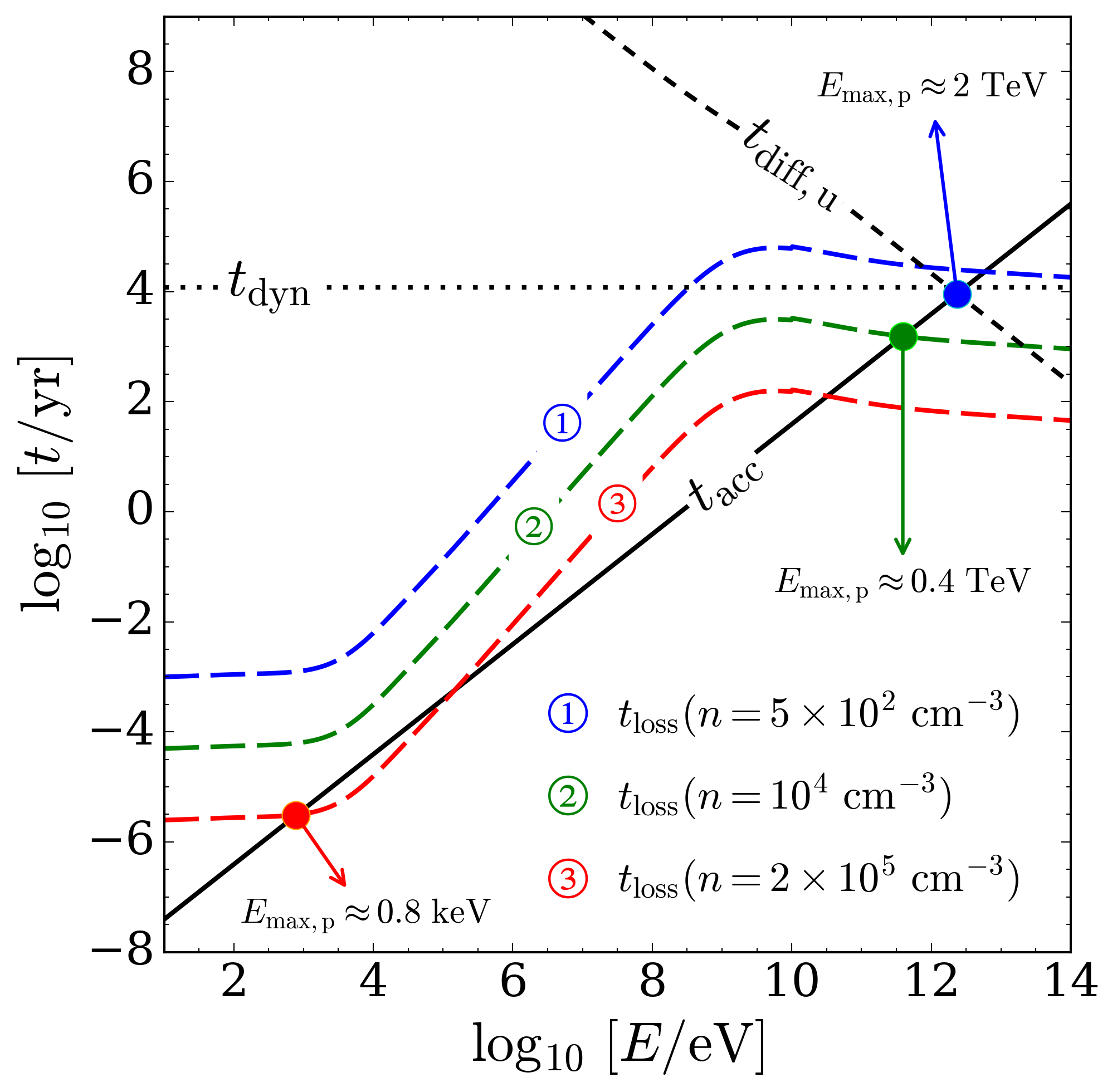}} 
\caption{Acceleration timescale ($t_{\rm acc}$, solid black line), upstream diffusion timescale ($t_{\rm diff,u}$, short-dashed black line), and dynamical timescale ($t_{\rm dyn}$, dotted black line) versus proton energy for $T=10^4$~K, $U=40$~km~s$^{-1}$, $B=100~\mu$G, and $R=10^5$~AU. The collisional loss timescale ($t_{\rm loss}$, long-dashed line plus colour-coding in the plot legend) depends on the density. The blue, green, and red dots show the value of the proton maximum energy for $n=5\times10^2$~cm$^{-3}$, $10^4$~cm$^{-3}$, and $2\times10^5$~cm$^{-3}$, respectively. We note that $E_{\rm max,p}$ is obtained by the intersection of $t_{\rm acc}$ with $t_{\rm diff,u}$ for $n=5\times10^2$~cm$^{-3}$, while
for the other two cases at higher $n$ by the intersection of $t_{\rm acc}$
with $t_{\rm loss}$.
}
\label{plottimescales}
\end{center}
\end{figure}

In order to emphasise which mechanisms determine $E_{\rm max,p}$, in Fig.~\ref{plottimescales}, we show the timescales for $T=10^4$~K, $U=40$~km~s$^{-1}$, $B=100~\mu$G, $R=10^5$~AU, and three different values of the volume density. For $n=5\times10^2$~cm$^{-3}$, $E_{\rm max,p}$ is determined by the upstream diffusion, whilst $n=10^4$~cm$^{-3}$  is constrained by collisional losses at high energies because of pion production. The timescale for collisional losses is shorter at low energies because of Coulomb losses.
For example, Coulomb losses reduce $t_{\rm loss}$ by a factor of about 2, 5, and 250 at 0.1~GeV, 0.1~MeV, and 1 keV, respectively.
This causes a strong damping 
in the acceleration efficiency at high volume densities (see $n=2\times10^5$~cm$^{-3}$ in Fig.~\ref{plottimescales}, and Sect.~\ref{outcomes}).

The emerging flux at the shock surface is determined by the fraction of the ram pressure, $nm_pU^2$, which is transferred to 
the
accelerated thermal particles, $\widetilde P$. Following \citet{be99}, $\widetilde P$ is proportional to the shock efficiency, $\eta$, which is the fraction
of thermal plasma particles entering the acceleration process, and
is given by
\be\label{eqPcr}
\widetilde{P}=\eta r\left(\frac{c}{U}\right)^{2}\widetilde{p}_{\rm inj}^{\,a}%
\left(\frac{1-\widetilde{p}_{\rm inj}^{\,b_{1}}}{2r-5}+\frac{\widetilde{p}_{\rm max}^{\,b_{2}}-1}{r-4}\right)\,,
\ee
where 
$a=3/(r-1)$, $b_{1}=(2r-5)/(r-1)$, $b_{2}=(r-4)/(r-1)$, and $\widetilde{p}_{k}=p_{k}/(m_{\rm p}c)$ is the normalised momentum.
Subscripts $k={\rm inj, max}$ refer to the injection (or minimum) momentum of a particle able to cross the shock that enters the process of acceleration, and the maximum momentum reached by an accelerated particle, respectively.
The injection momentum is related to the thermal particle momentum
\citep{bg05} by
\be\label{pinj}
p_{\rm inj}=\lambda p_{\rm th}=\lambda m_{\rm p} c_{\rm s,d}\,,
\ee
where $c_{\rm s,d}$ is the sound speed in the downstream region, obtained by Eq.~(\ref{cs}) for the downstream temperature. The parameter $\lambda$ is related to $\eta$ by
\be\label{lambdaeta}
\eta=\frac{4}{3\sqrt{\pi}}(r-1)\lambda^3{\rm e}^{-\lambda^2}\,.
\ee
While in supernova remnants $\widetilde P$ is assumed to be of the order of 10\%, shocks in \HII regions are slower 
and we expect \mbox{$\widetilde P <$ 10\%} since pressure is
smaller for slower shocks.
At the same time $\widetilde P$ has to be high enough in order to explain the synchrotron radiation flux. 
In the following we used 
a typical value of  $\widetilde P=1\%$, but we provide more accurate estimates for the Sgr B2(DS) case in Sect. \ref{applications}. 
We recursively computed $\eta$ by coupling Eqs.~(\ref{eqPcr}) and (\ref{lambdaeta}) with the further condition $\lambda\gtrsim2$, which guarantees that the accelerated particles are injected a few times $p_{\rm th}$.

\begin{figure}[!t]
\begin{center}
\resizebox{\hsize}{!}{\includegraphics{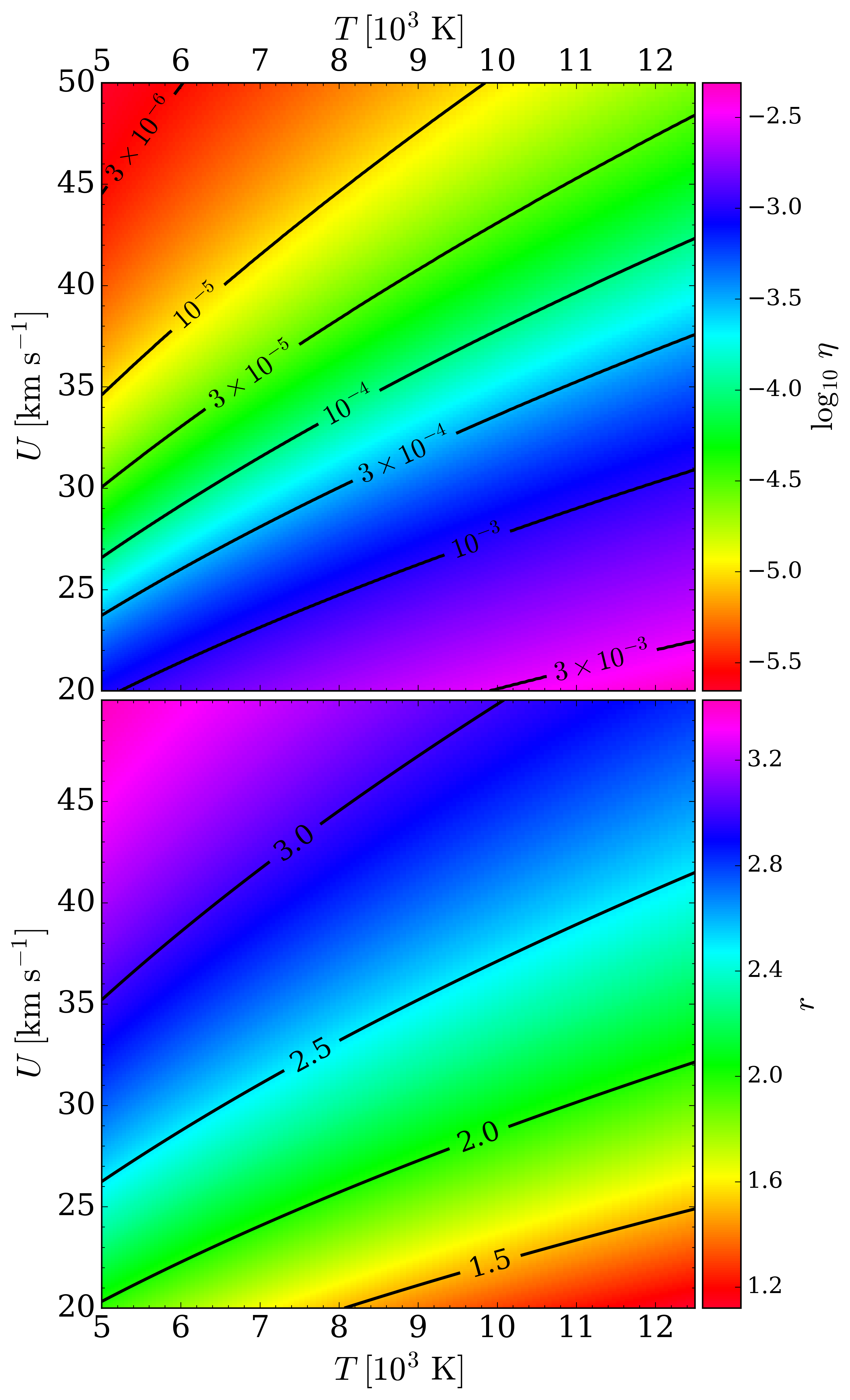}} 
\caption{Shock efficiency, $\eta$ (upper panel),
and compression ratio, $r$ (lower panel), 
for $\widetilde P=1\%$ 
as  function of shock temperature
and velocity. Solid black lines show iso-contours of values of 
$\eta$ and $r$.}
\label{etafig}
\end{center}
\end{figure}

Once $\widetilde P$ is fixed, $\eta$ depends only on the flow velocity 
in the shock reference frame and the temperature through $p_{\rm inj}$.
The upper panel of Fig.~\ref{etafig} shows $\eta$ for $\widetilde P=1\%$ and typical values of $T$ and $U$ expected in \HII regions 
(see Sect.~\ref{outcomes} and~\ref{applications}).
The lower panel shows that the compression ratios are always lower than four and the strong shock approximation never holds.\footnote{In the case of strong shocks,
the sonic Mach number is much larger than one (see Eq.~\ref{Mson}). As a result, $r\rightarrow4$ for $\gamma_{\rm ad}=5/3$, (see Eq.~\ref{compressionratio}).}

The energy distribution per unit density
(hereafter distribution)
of shock-accelerated protons is given by
\be\label{Nshp}
\mathscr{N}_p(E)=4\pi p^{2}f(p) \frac{\ud p}{\ud E}\,,
\ee
where $f(p)$ is the momentum distribution at the shock surface.
In the test-particle regime, 
the latter is described by a power-law momentum function 
\be\label{fp}
f(p)=f_{0}\left(\frac{p}{p_{\rm inj}}\right)^{-q}\,,
\ee
with $q=3r/(r-1)$. The normalisation constant, $f_0$, is given by
\be\label{f0}
f_0=\frac{3}{4\pi}\frac{n\widetilde P}{\mathscr{I}}\left(\frac{U}{c}\right)^2(m_pc)^{q-3}p_{\rm inj}^{-q}\,,
\ee
where
\be\label{I2}
{\mathscr I}=\int_{\widetilde{p}_{\rm inj}}^{\widetilde{p}_{\rm max}}\frac{p^{\,4-q}}{\sqrt{p^{\,2}+1}}\ud p\,.
\ee
%

\subsubsection{Acceleration of thermal electrons}\label{secEmaxe}
The electron injection process in shock acceleration is poorly understood.
As a guide, the model of \citet{bk00} is used to estimate 
the distribution 
of shock-accelerated electrons,
$\mathscr{N}_e$.
Taking the same energy of the injected protons
as that for electrons, namely $p_{\rm inj,e}=\sqrt{m_e/m_p}p_{\rm inj,p}$
($m_e$ is the electron mass), at relativistic energies it holds
\be\label{NeNp}
\frac{\mathscr N_e}{\mathscr N_p}=\left(\frac{m_e}{m_p}\right)^{(q-3)/2}\,.
\ee
The electron maximum energy, $E_{\rm max,e}$, is limited by synchrotron losses and it is
obtained by equating the acceleration timescale (Eq.~\ref{tacc}) to the synchrotron timescale,
$t_{\rm syn}$, which is given by
\be\label{tsyn}
t_{\rm syn}=2.7\times10^{11}%
\frac{\gamma-1}{\gamma^2}B_{-5}^{-2}~\mathrm{yr}\,.
\ee
If $t_{\rm syn}>t_{\rm acc}$ at any energy, then we set 
$E_{\rm max,e}=E_{\rm max,p}$. Finally, we also accounted for the fact
that at energies larger than $E^*$, where the condition
$t_{\rm syn}(E^*)<t_{\rm dyn}$ is fulfilled, 
the slope of the electron 
distribution, 
$s$, is modified from
$\mathscr{N}_e(E)\propto E^s$ to 
$\mathscr{N}_e(E)\propto E^{s-1}$ \citep{bg70}.

\begin{figure}[!t]
\begin{center}
\resizebox{\hsize}{!}{\includegraphics{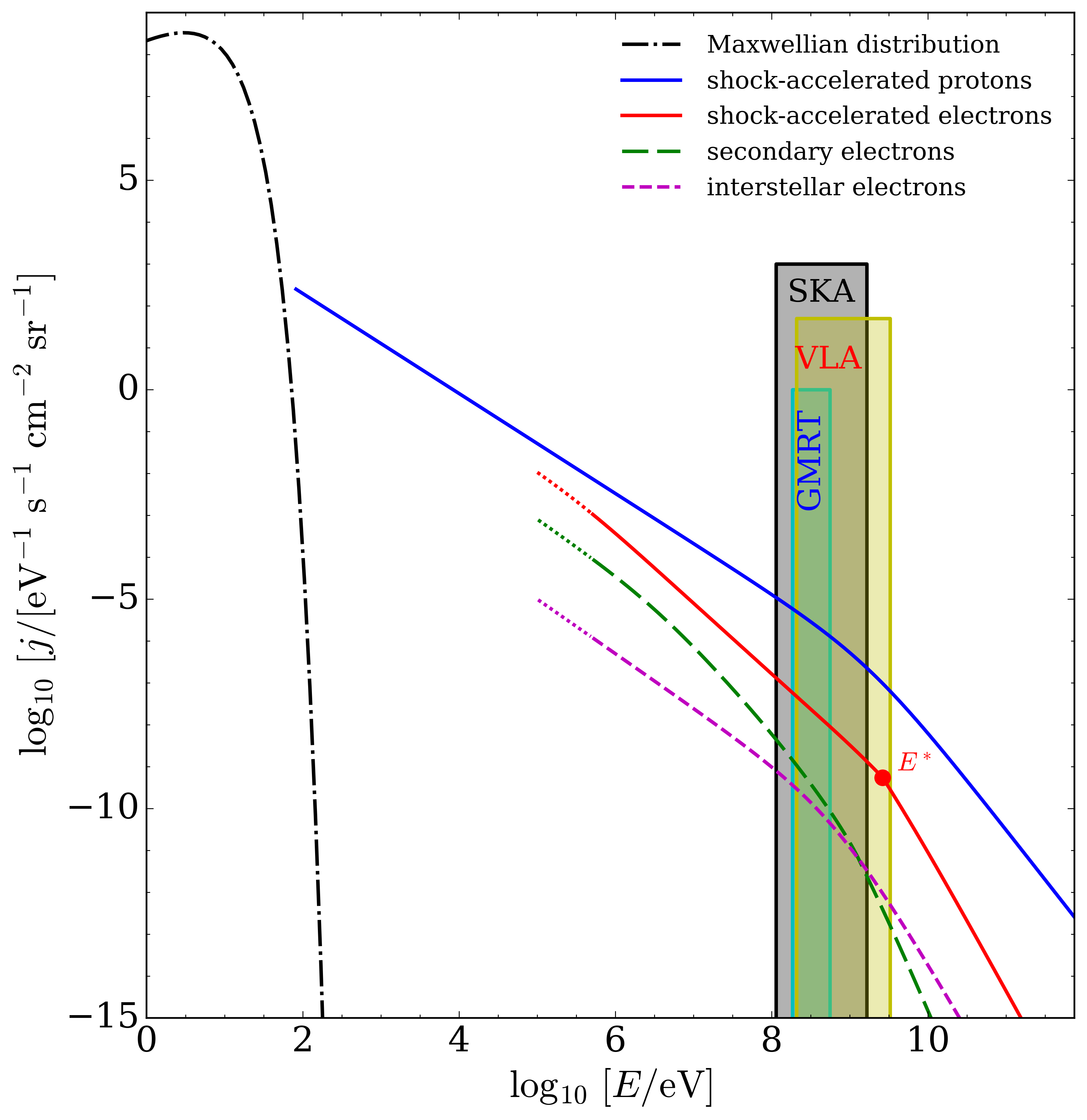}} 
\caption{Shock-accelerated fluxes of
protons (solid blue line) and
electrons (solid red line),
secondary electron flux (long-dashed green line), 
interstellar electron flux (short-dashed magenta line),
and Maxwellian distribution of thermal protons 
(dash-dotted black line)
as function of energy. The  solid red circle 
shows the energy
$E^*$ where synchrotron losses cause a break in the flux slope.
The cyan-, black-, and yellow-shaded areas show the electron energy
ranges mostly contributing to synchrotron emission for
$B=1$~mG and frequency ranges of GMRT, SKA, and 
VLA telescopes, respectively (see Eq.~\ref{Esyn}).
Dotted lines show the non-relativistic part of the electron fluxes.
}
\label{spectraexample}
\end{center}
\end{figure}

Figure~\ref{spectraexample} shows an example of proton and
electron fluxes emerging at the shock surface. Fluxes,
namely the number of particles per unit energy, time, area,
and solid angle,
were computed from the corresponding 
distributions 
as
\be\label{jfromN}
j_k=\frac{\beta_k c}{4\pi}\mathscr{N}_k\,,
\ee
where $k=p,e$.
Here we
consider $T=8\times10^3$~K, 
$U=50$~km~s$^{-1}$, $n=5\times10^4$~cm$^{-3}$,
$\widetilde P=1\%$, and $R=5\times10^4$~AU.
We set $B=1$~mG, high enough to shorten the synchrotron 
timescale so that the effect of the break in the slope of
$\mathscr{N}_e$
occurs (in this case at $E^*\simeq3$~GeV).

Most of the synchrotron radiation is emitted by electrons
of energy 
\be\label{Esyn}
E_{\rm syn}\approx1.47\left(\frac{\nu}{\rm GHz}\right)^{1/2}%
\left(\frac{B}{100~\mu{\rm G}}\right)^{-1/2}~\mathrm{GeV}\,,
\ee
see \citet{longbook} and \citet{pg18}.
Figure~\ref{spectraexample} shows the electron energy range
($E_{\rm syn}\approx0.1-5$~GeV)
that mostly contributes to synchrotron radiation for $B=1$~mG
and the three different frequency coverages of 
the Square Kilometre Array (SKA; $\nu=0.06-12.53$~GHz),
GMRT ($\nu=0.16-1.42$~GHz),
and VLA ($\nu=0.22-50$~GHz).

In this plot we also show the flux of secondary electrons
(computed following Appendix~B in
\citealt{ip15})
generated by shock-accelerated protons and electrons 
through ionisation losses as
soon as they propagate through 
a column density $N=10^{19}$~cm$^{-2}$,
corresponding to a 
distance of about 13~AU at $n=4\times10^5$~cm$^{-3}$.
This secondary electron flux is much smaller than the 
shock-accelerated
electron flux. 
For example, at 0.1, 1, and 5 GeV,
the ratio between the fluxes of shock-accelerated electrons
and secondary
electrons is about 30, 220, and 2900, respectively.
As a result, the contribution of secondary electrons
to the synchrotron flux 
density is negligible as well as that of the interstellar
electron flux obtained by the most recent Voyager~1 data release
\citep{cs16}. We note that Fig.~\ref{spectraexample}
shows the unattenuated interstellar electron flux. This has
to be considered as an upper limit since we expect strong attenuation effects at typical
volume densities of \HII regions \citep[see e.g.][]{pgg09,pi18}.

\section{Synchrotron emission in \HII regions}
\label{synchtrotonemission}
In this section we present the main equations 
for evaluating
the synchrotron flux density and its spectral index 
(Sect.~\ref{basiceqs}), which is followed by a description of the 
outcomes of the model applied to \HII regions (Sect.~\ref{outcomes}).\ Also, we
comment on the linearly polarised nature of synchrotron 
radiation (Sect.~\ref{polobs}).

\subsection{Basic equations}
\label{basiceqs}
The equations for computing flux density are listed in
the following. For a detailed review, see \citet{pg18}.
The total power per unit frequency emitted by an electron of energy $E$ at frequency $\nu$ is given by
\be\label{power}
P_{\nu}^{\rm em}(E)=\frac{\sqrt{3}e^{3}}{m_{e}c^{2}} B_{\perp} F\left[\frac{\nu}{\nu_{c}(B_{\perp},E)}\right]
\ee
%
(see e.g. \citealt{longbook}).
Here, $B_{\perp}$ is the projection of the magnetic field
on the plane perpendicular to the line of sight.
The function $F$ is defined by
\be
F(x)=x\int_{x}^{\infty}K_{5/3}(\xi)\ud\xi\,,
\ee
where $K_{5/3}$ is the modified Bessel function of order 5/3 and
\be\label{nuc}
\nu_{c}(B_{\perp},E)=
\frac{3eB_{\perp}}{4\pi m_{e}c}\left(\frac{E}{m_e c^2}\right)^{2}=
4.19\left(\frac{B_{\perp}}{\rm G}\right)%
\left(\frac{E}{m_e c^2}\right)^{2}~{\rm MHz}
\ee
is the frequency at which $F$ reaches its maximum value.
The synchrotron specific emissivity $\epsilon_{\nu}$ 
at frequency $\nu$,
namely the power per unit solid
angle and frequency produced within unit volume, is
\be\label{epsnu}
\epsilon_{\nu} = \int_{m_{e}c^{2}}^{\infty}\frac{j_{e}(E)}{\beta_e(E)c}P_{\nu}^{\rm em}(E)\,\ud E\,.
\ee
In principle, synchrotron self-absorption can take place
if synchrotron radiation is sufficiently 
strong. As a consequence, the emitting electrons absorb
synchrotron photons and the emission is quenched at low
frequencies \citep[see e.g.][]{rl86}. 
We computed the absorption coefficient per unit length 
as a function of the
frequency, $\kappa_\nu$, given by
\be\label{knu}
\kappa_{\nu}=-\frac{c^{2}}{2\nu^{2}}\int_{0}^{\infty}E^{2}\frac{\partial}{\partial E}%
\left[\frac{j_{e}(E)}{E^{2}\beta_{e}(E)c}\right]P_{\nu}^{\rm em}(E)\,\ud E
\ee
and the optical depth $\tau_\nu=\kappa_\nu L$, where $L$
is the dimension of the emitting region. 
The size of \HII regions changes with time and it 
can be related as a first approximation to the evolutionary stage \citep[see also][]{peters2010a, peters2010b}. The earliest stages are classified as
hypercompact, ultracompact, and compact \HII regions with 
sizes smaller than about 0.03~pc, 0.1~pc, and
0.5~pc, respectively \citep[see e.g.][]{ku05}.
Even assuming the largest size for $L$ and frequencies as low as 60~MHz (the lowest SKA1-low frequency\footnote{\url{https://astronomers.skatelescope.org/wp-content/uploads/2017/10/SKA-TEL-SKO-0000818-01_SKA1_Science_Perform.pdf}}), we find the expected synchrotron emission to always be optically thin ($\tau_\nu\ll1$). Thus, assuming a Gaussian beam profile, the synchrotron flux density at a frequency $\nu$, $S_\nu$, is given by
\be\label{Snu}
S_\nu=\frac{\pi}{4\ln2}\epsilon_\nu ~\theta_b^2~L\,,
\ee
where $\theta_b$
is the beam full width at half maximum.

\subsection{Model results}
\label{outcomes}
The main parameters regulating the non-thermal emission
in a \HII region are the temperature, the flow velocity
in the shock reference frame,
the magnetic field strength, the shock radius, and the dimension 
of the emitting region. In this section we computed the 
flux density assuming the following ranges which were obtained from observations and numerical simulations when possible:
$5\times10^3\leq T/{\rm K}\leq1.2\times10^4$, 
$20\leq U/({\rm km~s^{-1}})\leq100$,
$3\leq B/\mu{\rm G}\leq10^4$, and
$10^2\leq n/{\rm cm^{-3}}\leq 10^5$ 
\citep[see e.g.][]{zp67, heiles1981, wink1983, lockman1989, mehringer1993, sewilo2004, ku05, hs2011, sh17}.
For simplicity, we assumed the shock radius to be equal
to the dimension of the emitting region. Besides, since 
synchrotron 
emission has not been observed in hypercompact \HII 
regions so far, we let $L$ vary between 0.1 and 0.5~pc.

\begin{figure*}[!h]
\begin{center}
\resizebox{\hsize}{!}{\includegraphics{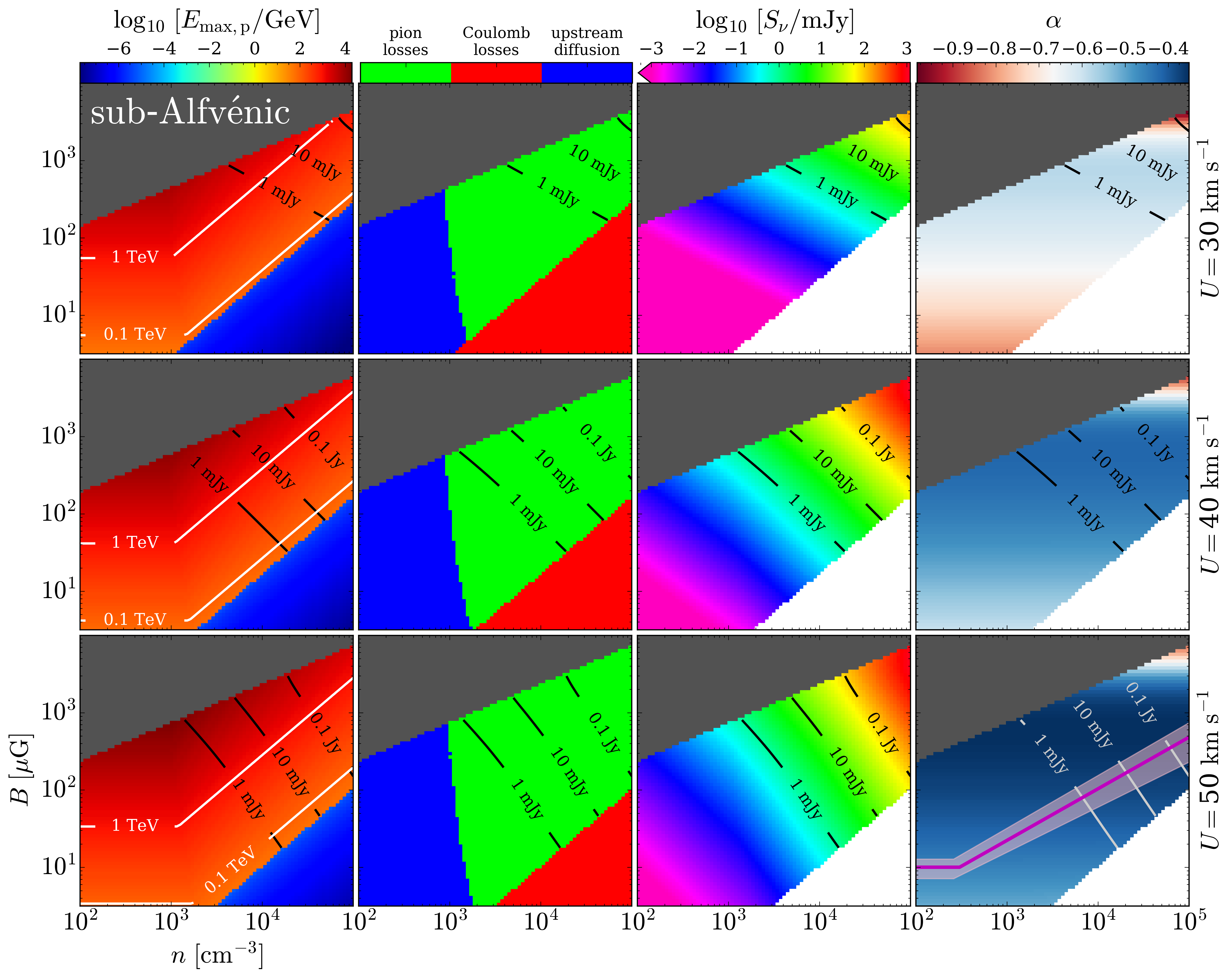}}
\caption{Model results for $T=7500$~K, $L=0.5$~pc, $\nu=300$~MHz, and
$\theta_b=10\arcsec$ in parameter
space $(n,B)$. 
Maximum energy of
shock-accelerated protons ($E_{\rm max,p}$, first column) and
its constraining timescales (second column), flux density 
($S_\nu$, third column), and spectral index ($\alpha$, fourth column).
The three rows show the above quantities at three different
flow velocities in the shock reference frame 
($U=30$, 40, and 50~km~s$^{-1}$).
Grey-shaded areas in each subplot show the region of the 
parameter space where the flow is sub-Alfv\'enic.
Solid white lines in the first column show the iso-contours
of $E_{\rm max,p}$ at 0.1 and 1~TeV. Solid black lines in
each subplot 
(except for the bottom right plot where black has
been replaced by light grey)
show the iso-flux density between 1 and 100~mJy
(from the bottom up).
The relation between magnetic field strength and density by
\citet{cru12} together with its uncertainty
is shown in the rightmost subplot of the lowest row
by a solid magenta line surrounded by a shaded region.
}
\label{results7500_300MHz_05pc_10as}
\end{center}
\end{figure*}

Figure~\ref{results7500_300MHz_05pc_10as} shows the results of 
our model for $T=7500$~K, $L=0.5$~pc, $\nu=300$~MHz, and
$\theta_b=10\arcsec$. 
Each subplot shows 
the various quantities in the parameter space
$(n,B)$. The three rows refer to three different
values of the velocities ($U=30$, 40, and $50$~km~s$^{-1}$) while
the four columns display the maximum energy reached by the
shock-accelerated protons, 
the mechanisms limiting $E_{\rm max,p}$,
the flux density, and the spectral index, respectively.
We note that at any density, for large magnetic field strengths,
the condition required by Eq.~(\ref{maxcsva}) is not always
fulfilled. In particular, the flow is sub-Alfv\'enic 
and no acceleration takes place (grey shaded area).

The iso-contours of $E_{\rm max,p}$ at 0.1 and 1~TeV in the first
column exhibit a kink around $1-2\times10^3$~cm$^{-3}$, which is
due to the variation in the mechanism controlling $E_{\rm max,p}$.
It is important to note that 
until the upstream diffusion constrains $E_{\rm max,p}$,
the iso-contours are independent on density since 
$t_{\rm diff,u}\not\propto n$ (Eq.~\ref{tescu}), 
but as soon as pion production losses
take place, $E_{\rm max,p}$ decreases with increasing density
since $t_{\rm loss}\propto n^{-1}$ (Eq.~\ref{tloss}).
As anticipated in Fig.~\ref{plottimescales},
the low-energy `tail' of
the collisional energy loss timescale determined by Coulomb
losses intersects the acceleration
timescale at high densities, thus 
$E_{\rm max,p}$ suddenly drops to non-relativistic
values 
and the acceleration process becomes ineffective.

Higher velocities increase the efficiency of shock 
acceleration and this affects the flux density magnitude, 
which increases
with $U$, as shown in the third column. 
Since $t_{\rm acc}\propto U^{-2}$, the solution space 
shrinks for lower velocities, namely Coulomb losses become progressively more efficient, as seen in the
plot of $S_\nu$ for $U=30$~km~s$^{-1}$.
The flux density increases with density
until Coulomb losses prevail because
$\mathscr{N}_p\propto f_0\propto n$ (Eqs.~\ref{Nshp}-\ref{f0}).
In addition, $S_\nu$ increases with magnetic field strength
until the flow becomes sub-Alfv\'enic
since the synchrotron emissivity (Eq.~\ref{epsnu}) is
proportional to $B^\delta$ with $s=1-2\delta$,
where $s$ is the slope of the electron distribution 
(see Sect.~\ref{secEmaxe} and \citealt{rl86}).

The rightmost column shows the spectral index, $\alpha$, 
of the flux density, $S_\nu\propto\nu^\alpha$ 
with $s=1+2\alpha$ 
\citep[see e.g.][]{rl86}.
The spectral index is independent of $n$, which only enters
in the normalisation factor $f_0$ (Eq.~\ref{f0}), while it
depends non-monotonically on $B$.
In fact, for small magnetic field strength, 
$E_{\rm syn}$ (calculated from Eq.~\ref{Esyn}) is larger than
the proton rest mass energy, 
where the proton (and electron) 
distribution slope is more negative than at
non-relativistic energies. As $B$ increases, $E_{\rm syn}$
moves towards non-relativistic energies so that $s$ 
(and $\alpha$) increases. Finally, at very large $B$, 
we may enter the regime where 
the synchrotron timescale (Eq.~\ref{tsyn}) is smaller
than the dynamical timescale (Eq.~\ref{tdyn}) 
and $\alpha$ decreases again
because of the slope break at $E^*$ (see Sect.~\ref{secEmaxe}). 
The rightmost plot in the lower row also shows the relation
between magnetic field strength and density, as given by
\citet{cru12}, that falls inside the solution space of our
model.

The dependence of $S_\nu$ on temperature in the range
considered in this paper is negligible. In fact, 
temperature enters 
the Coulomb part of the loss timescale 
(see \citealt{ms94} and Eq.~\ref{tloss}) so that  
the region of the parameter space
$(n,B)$ dominated by Coulomb losses becomes slightly
smaller for higher temperatures.
The injection momentum is also a function of the 
(downstream) temperature
through the downstream sound speed, 
$c_{\rm s,d}\propto\sqrt{T_{\rm d}}$
(Eq.~\ref{pinj}). However,
for increasing temperatures, $\eta$ increases
(see Fig.~\ref{etafig}) while $\lambda$ decreases
(Eq.~\ref{lambdaeta}), and $p_{\rm inj}$ turns out
to be only weakly dependent on temperature.
For example, $p_{\rm inj}$ varies by less than 2\%
between $T=5\times10^3$~K and $T=1.2\times10^4$~K
for $U=50$~km~s$^{-1}$.
Figure~\ref{results7500_300MHz_05pc_10as} has been
obtained for $\nu=300$~MHz and $\theta_b=10\arcsec$.
Since $S_\nu\propto\nu^{\alpha}$,
we expect lower flux densities at higher frequencies
since $\alpha$ is negative for non-thermal emission for a fixed beam size.


To facilitate the application of our model to a generic \HII
region, we developed a publicly available web-based application%
\footnote{\href{https://synchrotron-hiiregions.herokuapp.com}{\tt{https://synchrotron-hiiregions.herokuapp.com}}} that allows the user to compute 
the flux density per beam squared and per size of the emitting region,
$S_\nu/(\theta_b^2L)$, the spectral index, and the shock-accelerated
electron flux for a given set of parameters (temperature, shock
velocity, and observing frequency) in the parameter space ($n,B$).

\subsection{Polarisation observations}
\label{polobs}
Synchrotron emission is linearly
polarised \citep[see e.g.][]{longbook} and
the fractional polarisation, $\Pi$, is related to the electron
distribution slope and the spectral index by
\be\label{PI}
\Pi=\frac{1-s}{7/3-s}=\frac{3-3\alpha}{5-3\alpha}\,.
\ee
For example, spectral indexes between $-1$ and $-0.2$ correspond to
a fractional polarisation equal to 75\% and 64\%, respectively.
As a result, polarisation observations could be very useful 
to confirm the non-thermal nature of this emission.
These kind of observations would also be helpful in determining the type
of shock and whether it is parallel or perpendicular.\ It is important to note that there would be consequences on
the maximum energy reached by the accelerated particles
(see Sect.~\ref{timescales}).
In addition, information on the magnetic field morphology and the shock
type would allow us to refine the model and to describe the
propagation of shock-accelerated electrons (and protons) along the
magnetic field lines as soon as they leave the shock 
\citep[see e.g.][]{pg11,ph13}.

We note, however, that we assume
Bohm diffusion, namely the turbulent component of 
the magnetic field is of the order of the magnetic
field strength itself (see Sect.~\ref{nonbohm}).
As a consequence the magnetic field should be
randomly oriented, at least downstream, resulting
in no polarisation detection.
In the case of non-Bohm diffusion, the magnetic field 
is more ordered and the
detection of linear polarisation
could be more likely.
This could be the case for Sgr B2(DS) for
which we estimated $k_{\rm u}$ of the order
of 10 (see Sect.~\ref{applications}).
However, a high electron density at the shock position could cause Faraday depolarisation of the synchrotron emission, which is stronger at lower frequencies.




\section{Comparison with observations}
\label{applications}

There are some important caveats that we have to consider,
both when attempting to model a synchrotron source
as well as when interpreting observations.
The flux density that we obtained from observations is
the sum of the specific emissivity of each position along the
line of sight convolved with the beam. 
In fact, the magnetic field strength
and the locally accelerated
electron flux, which
determine the power per unit frequency
(Eq.~\ref{power}) and the emissivity (Eq.~\ref{epsnu}),
are `local' functions that namely depend
on position. 
Besides, in Sect.~\ref{secEmaxe} we show that the 
energy slope of shock-accelerated electrons is not
constant. It becomes more negative
above the proton
rest mass energy and there can be a further steepening
at energies larger than $E^*$ where the condition 
$t_{\rm syn}(E^*)<t_{\rm dyn}$ is satisfied
(see Fig.~\ref{spectraexample}).
For a given observing frequency, the energy of 
the electrons responsible for synchrotron emission
depends on the magnetic field strength
(Eq.~\ref{Esyn}), which is not 
constant along the line of sight. This means that 
we map different parts of the electron flux,
and different slopes,
as a function of the position, both along the line of sight
and on the plane of the sky. 

In our model we assume a constant
value for $B$ and a unique energy distribution of
shock-accelerated electrons. With only the knowledge of 
the volume
density profile and the magnetic field topology,
it would be possible to compute
the attenuation of the electron flux, 
accounting for magnetic effects on
particle propagation
(following e.g. \citealt{pgg09,pg11,ph13,pi18}).
As a result, by comparing modelled and observed flux
densities, we estimated an average value of
quantities such as temperature, flow velocity in the shock reference frame,
magnetic field strength, and volume density.
For a proper modelling of the flux density, one should 
know both the density and the magnetic field strength
profiles, as shown by \citet{pg18}.

Another important point is the way in which we interpret the
spectral index, $\alpha$.
The value of $\alpha$ estimated from observations 
can be strongly dependent 
on the observed frequencies. For example,
\citet{vv16} carried out GMRT observations at 325, 
610, and 1372~MHz in the \HII region IRAS 17256-3631, 
finding non-thermal emission at two positions.
The spectral indexes computed at lower frequency 
($325-610$~MHz) is clearly non-thermal
($\alpha=-0.91$ and $-1.25$), but at higher frequency
($610-1372$~MHz) the thermal contribution dominates
($\alpha=-0.07$ and $0.09$).
The benefit of
the model presented in this paper is that 
$\alpha$ is computed
as a local derivative of the flux density and this
allows us to make predictions on its value
at any frequency.

\citet{ms19} carried out
VLA observations at $4-12$~GHz towards the Sagittarius B2 complex,
finding a mixture of thermal and 
non-thermal emission in the `deep south'
region, 
hereafter Sgr~B2(DS).\ This is possibly due to an expanding \HII region.
This region has the shape of a shell with inner and outer radius
$R_{\rm in}\simeq0.36$~pc and $R_{\rm out}\simeq0.72$~pc,
respectively,
to which corresponds an 
average size of the emitting region 
$L=(\pi/2)R_{\rm out}(1-R_{\rm in}^2/R_{\rm out}^2)=0.85$~pc. 
Within this framework we applied the model described in
Sect.~\ref{model} in order to explain the origin of
the synchrotron emission. We assumed 
a parallel shock,
a temperature of $8\times10^3$~K (\citealt{mehringer1993};
\citealt{ms19}),
and the same beam size and
frequency range of VLA observations, namely 
$\theta_b=4\arcsec$ and $\nu=4-12$~GHz, respectively.
We also assumed $\widetilde P=5\%$ to
explain the non-thermal flux densities observed in DS.

For each of the five positions where a negative spectral index was computed from observations
(see Fig.~5 in \citealt{ms19}),
we 
compiled a 
library of models varying
flow velocities in the shock reference frame, densities,
and magnetic field strengths in the range 
$20\le U/({\rm km~s^{-1}})\le100$, $10^3\le n/{\rm cm^{-3}}\le10^5$, and $0.1\le B/{\rm mG}\le10$, respectively.
We note that the velocity range considered
is in agreement with what was obtained by simulations of cometary \HII regions of O and B stars
driving strong stellar winds \citep{sh17}.
We 
performed a $\chi^2$ test
identifying the best $U$, $n$,
and $B$ values that
reproduce the observed flux densities.
At first we considered the Bohm diffusion
regime ($k_{\rm u}=1$) then, using the values of $U$, $n$, and $B$
from the $\chi^2$ test, we recomputed the upstream
diffusion coefficient following \cite{pl06}
\be
k_{\rm u}=4\times10^{-4}U_2^{-1}n_6^{-0.5}B_{-5}
\widetilde{P}^{-1}\,,
\ee
and repeated the procedure till
$k_{\rm u}$ converges. We found $k_{\rm u}$ to be of the order of ten
for all five positions, which indicates a regime of non-Bohm
diffusion.

\begin{figure*}[!ht]
\begin{center}
\resizebox{\hsize}{!}{\includegraphics{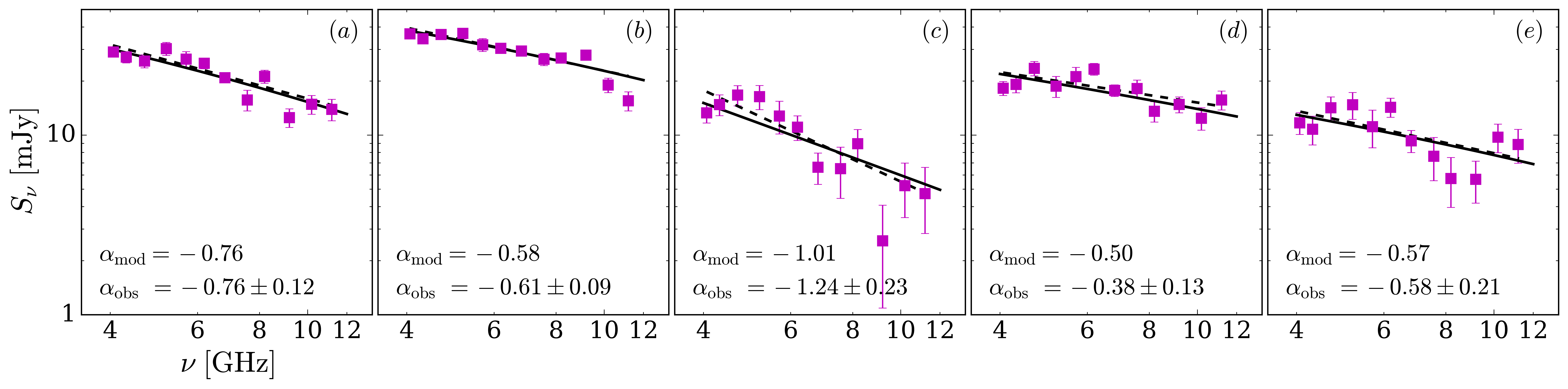}}
\caption{Observed flux densities (magenta squares) and their best fits (dashed black lines) for five positions in DS
as function of frequency (labelled $(a)$ to $(e)$; 
see Fig.~5 in \citealt{ms19}).
Solid black lines show the model results (see Tab.~\ref{tab:comparison} for a complete overview of 
the model parameters). Each subplot also displays the modelled and
observed spectral indexes, $\alpha_{\rm mod}$
and $\alpha_{\rm obs}$, respectively.}
\label{SgrB2}
\end{center}
\end{figure*}

The results of the $\chi^2$ minimisation are shown in Table~\ref{tab:comparison}, where the $1\sigma$
errors on $U$, $n$,
and $B$ were estimated
using the method of \citet{lm76}.
The observed flux densities fall in the range
$1-40$~mJy and they were reproduced by
our model with an average accuracy of 5\%
for magnetic field strengths spanning between
about 0.6 and 3~mG, 
densities between about 3 and $6\times10^4$~cm$^{-3}$, and
shock velocities between 35 and 45~km~s$^{-1}$
(see Fig.~\ref{SgrB2}).
We noticed
that our model allowed us to discard velocities lower than about
35~km~s$^{-1}$, which
cannot explain 
the observed flux densities since
$t_{\rm acc}\propto U^{-2}$ and
the particle acceleration process becomes
rapidly inefficient for decreasing velocities
(see Sect.~\ref{outcomes}).
It is remarkable that the modelled spectral 
indexes, $\alpha_{\rm mod}$, which are obtained as a by-product of the 
$\chi^2$ minimisation, are also within the error bars of the observed
spectral indexes 
$\alpha_{\rm obs}$ 
(see 
Table~\ref{tab:comparison}). 
It is interesting to note that in \citet{ms19},
we applied the model
to explain the non-thermal
emission in the whole Sgr B2(DS) region with an average accuracy lower 
than 20\%. We obtained the distributions of velocities, densities,
and magnetic field strengths, which give information on the dynamics of DS.
In fact, we found low velocities ($33\le U/({\rm km~s^{-1}})\le40)$  
towards north, where 
density and magnetic field strength are higher, and
velocities in the range $40-50$~km~s$^{-1}$ in the transverse east-west
direction, where density and magnetic field strength are lower, as if the 
\HII region were expanding towards the direction of minimum resistance
(see \citealt{ms19} for details).


\begin{table*}[!h]
    \centering
    \begin{tabular}{ccccccc}
    \hline\hline
    position & $U$ & $n$ & $B$ & $\langle (S_{\nu,\rm obs}-S_{\nu,\rm mod})/S_{\nu,\rm obs}\rangle$ & $\alpha_{\rm mod}$ & $\alpha_{\rm obs}$ \\
             & [km s$^{-1}$] & [$10^4$~cm$^{-3}$] & [mG] & [\%] & & \\
    \hline\\[-0.25cm]
    $a$ & $44_{-0.4}^{+0.3}$ & $3.50_{-0.10}^{+0.05}$ & $1.44_{-0.11}^{+0.11} $ & 4.0 & $-0.76$ & $-0.76\pm0.12$\\[0.1cm]
    $b$ & $45_{-0.2}^{+0.2}$ & $4.55_{-0.03}^{+0.03}$ & $0.957_{-0.028}^{+0.031} $ & 1.1 & $-0.58$ & $-0.61\pm0.09$\\[0.1cm]
    $c$ & $34_{-0.2}^{+0.2}$ & $5.92_{-0.54}^{+0.48}$ & $3.26_{-0.09}^{+0.10} $ & 8.3 & $-1.01$ & $-1.24\pm0.23$\\[0.1cm]
    $d$ & $45_{-0.3}^{+0.6}$ & $4.17_{-0.06}^{+0.10}$ & $0.574_{-0.031}^{+0.026}$ & 6.0 & $-0.50$ & $-0.38\pm0.13$\\[0.1cm]
    $e$ & $43_{-1.1}^{+1.0}$ & $2.47_{-0.19}^{+0.19}$ & $0.864_{-0.010}^{+0.015} $ & 2.9 & $-0.57$ & $-0.58\pm0.21$\\[0.1cm] 
    \hline
    \end{tabular}
    \caption{Results of the $\chi^{2}$ minimisation for the five positions observed in Sgr~B2(DS),
    see Fig.~5 in \citet{ms19}.
    Flow velocity in the shock reference frame, volume
    density, and magnetic field strength (columns 2 to 4),
    average difference in percent between the observed and the modelled flux density
    ($S_{\nu,\rm mod}$ and $S_{\nu,\rm obs}$; column 5), and 
    modelled and observed spectral indexes 
    ($\alpha_{\rm mod}$ and $\alpha_{\rm obs}$; columns 6 and 7).}
    \label{tab:comparison}
\end{table*}

Sgr~B2(DS) is a special case where non-thermal emission is found
all along the ionised bubble, especially in the inner part, which is
expected to be closer to the shock and namely to the particle acceleration
site. As a result, the application of the model is more
straightforward
with respect to \HII regions, such as those associated with
IRAS~17160-3707 \citep{nv16} and IRAS~17256-3631 \citep{vv16}.
Indeed, these latter sources show thermal emission with some 
localised spots of non-thermal emission.\ Additionally, proper modelling
would require knowledge of the spatial 
variation of all the parameters including the flow velocity
in the shock reference frame,
density, and magnetic field strength.
This goes beyond the scope of the present paper, but we presume
that non-thermal emission spots in these IRAS sources could be 
explained by shock-accelerated electrons as in Sgr~B2(DS).




\section{Discussion and conclusions}
\label{conclusions}

We explored the possibility of thermal electrons 
accelerating up to
relativistic energies in \HII
regions through the first-order Fermi acceleration mechanism,
assuming that shocks are located at the 
position where non-thermal emission is detected through radio
observations. We computed the shock-accelerated electron
flux and the corresponding flux density by assuming a completely
ionised medium and by studying the parameter
space for the following quantities: 
temperature ($5-12\times10^3$~K),
flow velocity in the shock reference frame ($20-100$~km~s$^{-1}$),
magnetic field strength ($3-10^4~\mu$G),
and density ($10^2-10^5$~cm$^{-3}$).

We found that the modelled flux densities are of the same order
of those observed and we concluded that non-thermal emission
in \HII regions might be due to synchrotron radiation from locally
accelerated electrons braked in a magnetic field.
Assuming Bohm diffusion, this mechanism is efficient if 
\mbox{$U>30$~km~s$^{-1}$}. These high velocities
are of the same order of those obtained by numerical simulations of O and
B stars generating \HII regions by powerful stellar winds. 
The acceleration efficiency is
also quenched as soon as the medium is not completely 
ionised ($x<1$) because ions and neutrals are not coupled; 
even an ionisation degree of 95\% strongly reduces the range of
densities and magnetic field strengths corresponding to 
flux densities
comparable to those observed.

We applied our model to Sgr~B2(DS) in order to explain the observed
flux densities and the spectral indexes 
\citep{ms19}. 
By means of
a $\chi^2$ test, we found that flux densities in the frequency
range $4-12$~GHz can be carefully 
reproduced by our model with 
an average accuracy lower than 20\% by assuming a parallel shock, a
completely ionised medium, a temperature of 8000~K, and for magnetic field
strengths, densities, and velocities in the ranges
$0.3-4$~mG,
$1-9\times10^4$~cm$^{-3}$, 
and $33-50$~km~s$^{-1}$, respectively.
It is worth noting that the modelled spectral indexes, 
which are 
a by-product of the $\chi^2$ test, fall within the errors of the
spectral indexes
computed from observations. 
Considering the relative simplicity of our model, this is a
promising result, even if in principle one should model an \HII region 
accounting for the spatial 
variation of density, magnetic field strength, and velocity, for example.

The best model for Sgr~B2(DS)
is obtained for an upstream diffusion coefficient
of the order of ten, namely $\delta B\approx B/3$.
This means that the magnetic field is not
completely randomly oriented, such as in the Bohm
diffusion case.  
Future polarisation observations would be very useful 
$(i)$ to confirm the non-thermal origin of this emission,
since synchrotron
emission is highly linearly polarised, and
$(ii)$ to clarify 
the nature of 
shocks in \HII regions, in particular whether they are parallel, perpendicular, or oblique.
Information on magnetic field morphology would also allow us
to account more accurately for the propagation of locally-accelerated
electrons along magnetic field lines once they leave the shock.

We also developed an interactive on-line tool that allows a fast
application of our modelling without going 
through all the equations.
The tool computes the shock-accelerated electron flux, the flux density,
and the spectral index in the parameter space $(n,B)$
for a given set of temperatures, flow velocity in the shock reference frame, 
and observing frequency.

Higher sensitivity, larger field 
of view, higher survey speed, and polarisation
capability of future telescopes
such as SKA will allow us to find a larger
number of \HII regions associated with non-thermal 
emission, giving us the opportunity to 
better characterise
the origin of synchrotron emission in \HII
regions.


\begin{acknowledgements} 
The authors thank Harrison Steggles for sharing relevant information on velocity fields in \HII regions from his simulations and the referee, Luke Drury, for his careful reading of the manuscript and insightful comments.
MP acknowledges funding from the European Unions Horizon 2020 research and innovation programme under the Marie Sk\l{}odowska-Curie grant agreement No 664931. ASM, FM, and PS research is carried out within the Collaborative Research Centre 956, sub-project A6, funded by the Deutsche Forschungsgemeinschaft (DFG) --- project ID 184018867.
\end{acknowledgements} 

\bibliographystyle{aa} 
\bibliography{mybibliography.bib} 

\appendix

\section{Shock acceleration in an incomplete ionised \HII region}\label{app:xneq1}
\begin{figure}[!t]
\begin{center}
\resizebox{.8\hsize}{!}{\includegraphics{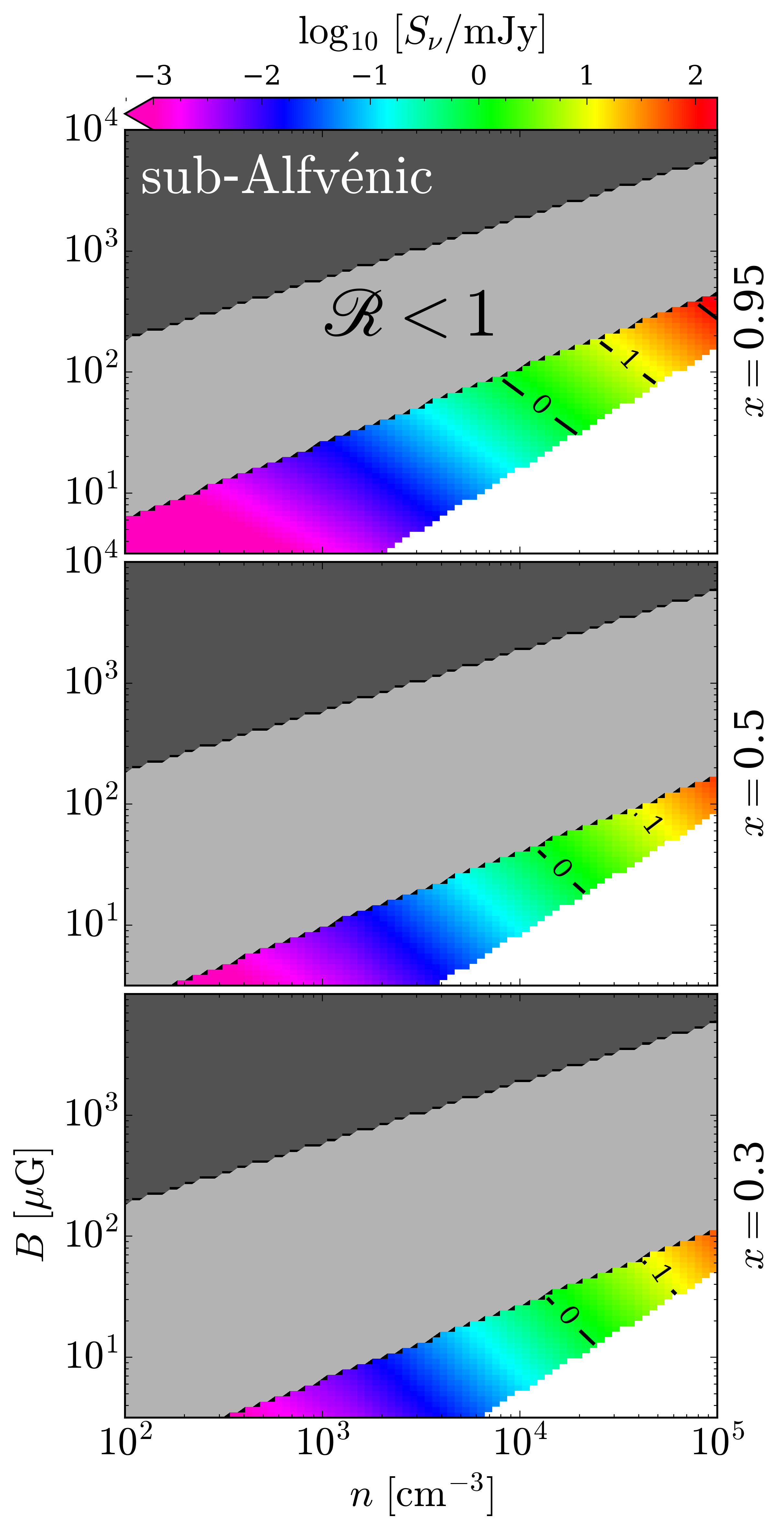}}
\caption{Flux density, $S_\nu$, for $T=7500$~K, $U=40$~km~s$^{-1}$,
$L=0.5$~pc, $\nu=300$~MHz, and
$\theta_b=10\arcsec$ in parameter
space $(n,B)$ for three different values of  ionisation fraction
as labelled on right-hand side of each subplot. 
Dark and light grey-shaded areas in each subplot show the region of the 
parameter space where the condition of 
super-Alfv\'enic flow and ion-neutral coupling is not satisfied,
respectively.
Solid black lines in
each subplot show the iso-flux density at 1 and 10~mJy in logarithmic
scale (from the bottom up).
}
\label{results7500_300MHz_05pc_10as_U40_x03_05_095}
\end{center}
\end{figure}

In our model we assumed that the medium is completely ionised. 
However, if $x<1$, the frictional force between ions and neutrals
can quench the acceleration process. The latter can still be 
efficient if charges and neutral particles are coupled so that 
ion-generated waves are weakly damped. The coupling condition is 
obtained by imposing
that the momentum transfer rate from ions to neutrals is larger than the wave pulsation \citep[see][]{dd96,ph15,pm16}.
Then, the upper energy limit due to the wave damping is found by equating the local accelerated particle flux advected downstream by the flow to the flux lost upstream because of the lack of waves to confine the accelerated particles due to
the wave damping (see \citealt{dd96} and Appendix~D in \citealt{pm16}).
The above conditions can be combined in a single relation \citep[see Eqs.~$8-9$ in][]{ph15}
\be\label{eqR}
\mathscr{R}=\frac{10^2}{\beta}\Xi U^3_2 n_6 x^{1.5}(1-x)^{-1} B_{-5}^{-6}\widetilde{P}\,,
\ee
where
\be\label{Xi}
\Xi=B^4_{-5}+1.4\times10^{12}\gamma^2\beta^2T_4^{0.8}n_6^3x^2\,.
\ee
If $\mathscr{R}<1$, charges and neutrals are not coupled anymore and the
acceleration mechanism is quenched.
Figure~\ref{results7500_300MHz_05pc_10as_U40_x03_05_095} shows the results
for an \HII region with $T=7500$~K and $L=0.5$~pc observed at $\nu=300$~MHz
with a beam $\theta_b=10\arcsec$ (such as in 
Fig.~\ref{results7500_300MHz_05pc_10as}), for a representative velocity of
40~km~s$^{-1}$, but considering an incomplete ionised medium.
In particular, we explored an ionisation fraction $x=0.95$, 0.5, 
and 0.3.
We find that even a small departure from $x=1$ causes a strong
reduction of the solution space. 

\end{document}